\documentclass[pra,aps,showpacs,floatfix,twocolumn,nofootinbib,superscriptaddress]{revtex4-1}
\usepackage{amsmath}
\usepackage{amsfonts}
\usepackage{amssymb}
\usepackage{revsymb}
\usepackage{graphicx}
\usepackage{physics}
\usepackage{xcolor} 
\allowdisplaybreaks

\def\ket#1{|#1\rangle}

\def\matrixelred#1#2#3{\langle #1\| #2\|#3\rangle}

\begin{document}
\title{Precision measurement of the $^3D_1$ and $^3D_2$ quadrupole moments in Lu$^+$}
\author{R. Kaewuam}
\affiliation{Centre for Quantum Technologies, National University of Singapore, 3 Science Drive 2, 117543 Singapore}
\author{T. R. Tan}
\affiliation{Centre for Quantum Technologies, National University of Singapore, 3 Science Drive 2, 117543 Singapore}
\affiliation{Department of Physics, National University of Singapore, 2 Science Drive 3, 117551 Singapore}
\author{Zhiqiang Zhang}
\affiliation{Centre for Quantum Technologies, National University of Singapore, 3 Science Drive 2, 117543 Singapore}
\author{K. J. Arnold}
\affiliation{Centre for Quantum Technologies, National University of Singapore, 3 Science Drive 2, 117543 Singapore}
\affiliation{Temasek Laboratories, National University of Singapore, 5A Engineering Drive 1, 117411 Singapore}
\author{M. S. Safronova}
\affiliation{Department of Physics and Astronomy, University of Delaware, Newark, Delaware 19716, USA}
\affiliation{Joint Quantum Institute, National Institute of Standards and Technology and the University of Maryland,
College Park, Maryland, 20742}
\author{M. D. Barrett}
\email{phybmd@nus.edu.sg}
\affiliation{Centre for Quantum Technologies, National University of Singapore, 3 Science Drive 2, 117543 Singapore}
\affiliation{Department of Physics, National University of Singapore, 2 Science Drive 3, 117551 Singapore}
\date{\today}
\begin{abstract}
Precision measurements of the Lu$^+$ $^3D_1$ and $^3D_2$ quadrupole moments have been carried out giving $\Theta(^3D_1)=0.63862(74)\,e a_0^2$ and $\Theta(^3D_2)=0.8602(14)\,e a_0^2$, respectively.  The measurements utilize the differential shift between ions in a multi-ion crystal so that effects of external field gradients do not contribute leaving only the well defined Coulomb interaction.  At this level of precision, hyperfine-mediated corrections will likely be important.
\end{abstract}
\maketitle
\section{Introduction}
The atomic electric quadrupole (EQ) moment is a consequence of an atomic state's non-spherical charge distribution, and results in an interaction between the atom and externally applied electric field gradients.  The resulting shift of the atomic energy level has been an important factor in the development of optical atomic clocks, both in terms of accurate assessment of the shift and methods to suppress it.  Measurement of the quadrupole moment entails the application of a well defined gradient field, and measuring the shift of the atomic level.  However this can be difficult:  the shift itself is typically at the Hertz level, which can make it difficult to measure against other effects, such as magnetic field noise, and there is an orientation dependence that can be influenced by stray fields and trap imperfections.

Measurements of the EQ moment have been achieved in a number of different ways.  Frequency shifts of optical clock transitions has been used in Hg$^+$ \cite{oskay2005measurement}, Sr$^+$ \cite{barwood2004measurement}, and Yb$^+$ \cite{schneider2005sub} with inaccuracies in the 4-12\% range.  Another approach has been to measure differential shifts between neighbouring Zeeman states, which eliminates many of the systematics that can influence optical measurements. This approach has achieved inaccuracies of $\sim 0.5\%$ using either dynamic decoupling \cite{shaniv2016atomic} or decoherence-free entangled states \cite{roos2006designer} to mitigate problems with magnetic field noise.  In all cases, the quadrupole shift is induced by the dc confinement fields including stray fields, which can have a non-negligible contribution \cite{dube2013evaluation}.  To avoid the influence of stray fields, another approach utilized the resonant coupling induced by the oscillating quadrupole field when the Zeeman splitting is half the frequency of the rf trap drive \cite{arnold2019oscillating}.  Although this method eliminated the dependence on stray fields, it was limited by decoherence from magnetic field noise.

Recently, an alternative strategy made use of the differential frequency shift between ions in a multi-ion crystal \cite{tan2019suppressing}.  In this case the shift is almost completely determined from the electric field gradient arising from neighbouring ions: confinement and stray fields appear common mode and magnetic field gradients are small and can be easily compensated.  The proof of principle demonstration in \cite{tan2019suppressing} was primarily limited by magnetic field alignment with respect to the crystal axis.  Here we provide an order of magnitude improvement over the previous measurements and also apply the technique to $^3D_2$.  We obtain a measurement precision that is statistically limited to the $\sim0.1\%$ level.

The measurements provide a high-precision benchmark test of the theoretical predictions of the quadrupole moments for Lu$^+$ \cite{porsev2018clock}, as well as the method to evaluate theoretical uncertainty. Previous such tests have been carried out for monovalent systems, such as Ca$^{+}$ \cite{roos2006designer} and Sr$^{+}$ \cite{shaniv2016atomic}. Theoretical calculations in Lu$^{+}$ are carried out by a different method that combines configuration interactions and coupled-cluster approaches (CI+all-order method) \cite{SafKozJoh09}. This method is used to predict atomic properties for a large variety of systems, including new clock proposals with highly-charged ions \cite{KozSafCre18}. To the best of our knowledge, only two high-precision benchmarks exists for any of the properties involving quadrupole matrix elements for systems with more than one valence electron \cite{PorSafKoz12,MasHoeWan19}, neither of which involve quadrupole moment measurements.  Present measurements directly demonstrate the validity of the theory and the corresponding evaluation of uncertainties in the calculated quadrupole moments, essential for guiding future experiments with highly charged ions and other systems \cite{PorSafSaf20}.

\section{Experimental System}
Experiments are carried out in a linear Paul trap with axial end caps as used in previous work\cite{tan2019suppressing,kaewuam2020hyperfine}.  The trap drive voltage of frequency $\Omega_\mathrm{rf}=2\pi\times 16.805\,\mathrm{MHz}$ is applied to diagonally opposing electrodes via a quarter wave resonator.  Relative to earlier work, the trap drive voltage is lowered and the dc end cap potentials raised, which reduces the amount of axial micromotion on the outer ions and increases the quadrupole shift between ions.   The resulting confinement frequencies for a single lutetium ion ($^{176}$Lu$^+$) are approximately $391(1)\,\mathrm{kHz}$ and $446(1)\,\mathrm{kHz}$ in the radial direction and $198.705(50)\,\mathrm{kHz}$ in the axial direction.  A magnetic field of $\sim 0.1\,\mathrm{mT}$ is used to lift the Zeeman degeneracy.  Fluorescence at 646\,nm is imaged onto an EMCCD camera with sufficient resolution to provide detection efficiencies of better than 99\% for each ion in a three-ion crystal.

\begin{figure*}
  \includegraphics[width=6.9in]{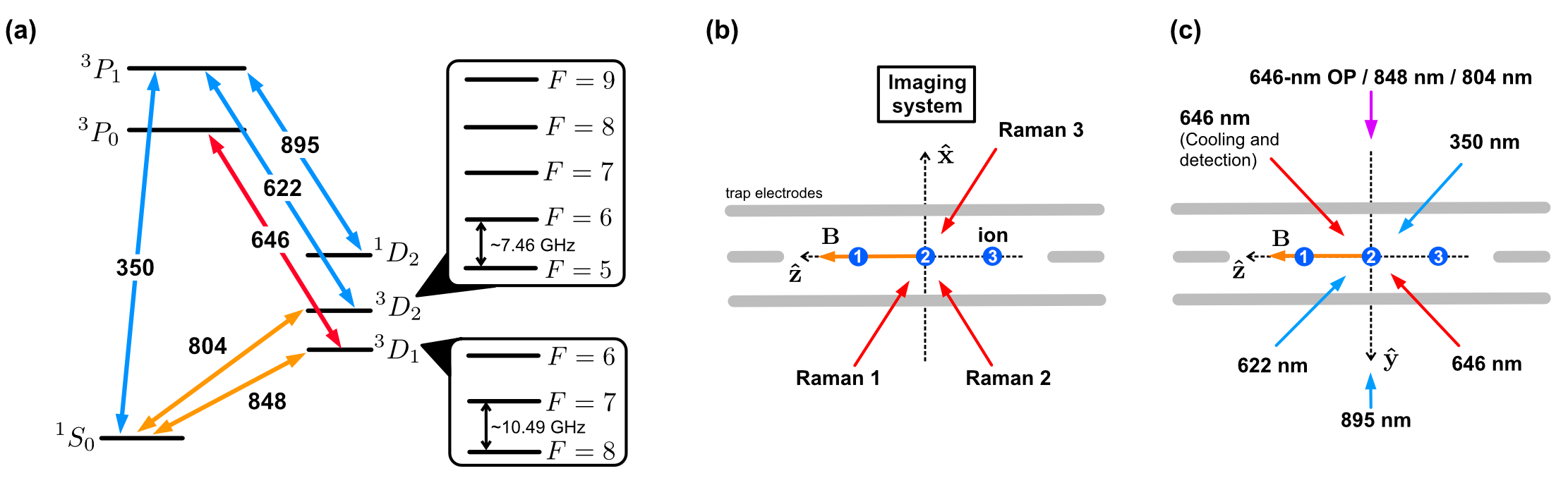}
  \caption{(a) Relevant energy level and transition diagram of a $^{176}$Lu$^+$. The wavelengths of the corresponding transitions are given in nm units. The two small boxes show the hyperfine structures of $^3D_1$ and $^3D_2$ with the microwave transitions used in this work. (b) and (c) show the schematic of the experimental setup of laser beam configurations relative to the trap: views along the $\vu{y}$ (vertical) and $\vu{x}$ (horizontal) axes, respectively. The applied dc magnetic field $\vb{B}$ is aligned along the trap axis maximizing the EQ shift.  The field direction is controlled by three Helmholtz coils. The vertical alignment is constrained by the 646-nm optical pumping (OP) beam (see text). The Raman beams in (b) and a 804-nm clock beam in (c) are used to assess EMM.}
 \label{ExpSetup}
\end{figure*}

Singly ionized lutetium has two valence electrons with a $^1S_0$ ground state and low lying $D$ states, which provide three optical clock transitions.  The $^3D_1$ and $^3D_2$ levels are the focus of this work and the relevant level structure is shown in Fig.~\ref{ExpSetup}(a).  Optical pumping with laser beams at 350, 622, and 895\,nm initializes population in the $^3D_1$ level.  Doppler cooling and fluorescence measurements are accomplished with laser beams at 646\,nm addressing the $^3D_1\leftrightarrow{}^3P_0$ transition.  An additional $\pi$-polarized 646-nm beam addressing $F=7$ to $F'=7$ facilitates state preparation into $\ket{{}^3D_1,7,0}$.  Raman beams at 646\,nm provide sideband spectroscopy of the $\ket{^3D_1,7,0}\leftrightarrow\ket{^3D_1,8,0}$ transition for the assessment of micromotion and measurement of trap frequencies.  The two lasers at 848 and 804\,nm drive the $^1S_0\leftrightarrow{}^3D_1$ and $^1S_0\leftrightarrow{}^3D_2$ clock transitions, respectively.  All laser beams are switched with acousto-optic modulators (AOM).  Laser configurations relative to the trap are as shown in Fig.~\ref{ExpSetup}.  Microwave transitions between hyperfine levels are driven by an antenna located outside the vacuum chamber.

The clock lasers at 848 and 804\,nm are extended-cavity-diode lasers (ECDL), which are phase locked via an optical frequency comb (OFC).  The short term ($<10\,$s) stability of the OFC is derived from the $\sim 1\,\mathrm{Hz}$ linewidth laser at 848 nm, which is referenced to a 10\,cm long ultra-low expansion (ULE) cavity with finesse of $\sim4\times10^5$. For longer times ($\gtrsim 10\,\mathrm{s}$), the OFC is steered to an active hydrogen maser (HM) reference.  Both clock lasers are $\pi$-polarized allowing $\Delta m=\pm1$ transitions to be driven on their respective clock transitions.  In addition, the phase lock between the two lasers enables a direct, two-photon transition from $\ket{{}^3D_1,7,0}$ to $\ket{{}^3D_2,5,0}$.  For this transfer, the clock lasers are detuned by several MHz from their respective clock transitions so they do not influence the transfer.

\section{Theory}
In a linear ion crystal, the EQ shift of $\ket{F,m_F}$ induced on the $i^\mathrm{th}$ ion due to all other ions is given by \cite{Itano2000,tan2019suppressing}
\begin{equation}
h\Delta f^{(Q)}_{F,i} =Q_i (3 \cos^2 \theta-1)\Theta(J,F,m_F) \frac{m \omega_z^2}{4e},
\label{QuadrupoleEq}
\end{equation}
where $e$ is the elementary charge, $m$ is the mass of the ion, $\omega_z$ axial confinement, $\theta$ is the angle between the applied dc magnetic field and the trap axis, and $Q_i$ is a scale factor dependent on the position of the position.  For a three-ion crystal, $\{Q_1,Q_2,Q_3\} = \{9,16,9\}/5$.  The EQ moment can be written $\Theta(J,F,m_F)=C_{F,m_F} \Theta(J)$ where 
\begin{multline}
C_{F,m_F}= (-1)^{2F+I+J-m_F}(2F+1)\\
\times \begin{pmatrix} F & 2 & F\\ -m_F & 0 & m_F\end{pmatrix}\begin{Bmatrix} F & F & 2\\ J & J & I\end{Bmatrix}\begin{pmatrix} J & 2 & J\\ -J & 0 & J\end{pmatrix}^{-1},
\end{multline}
and $\Theta(J)$ is the quadrupole moment for the fine-structure level defined by
\begin{equation}
\Theta(J)=\begin{pmatrix} J & 2 & J\\ -J & 0 & J\end{pmatrix}\matrixelred{J}{\Theta^{(2)}}{J}.
\end{equation}

A magnetic field gradient and excess micromotion (EMM) can also provide a small contribution to the differential shift between ions.  In a three ion crystal, a magnetic field gradient provides an equal and opposite shift on the two outer ions relative to the inner ion.  Although the gradient can be measured accurately in our system, the effects maybe cancelled by averaging over the two outer ions.  This leaves only the contribution from EMM.

The only contribution from EMM relevant to a microwave transition is that from the tensor component of the ac Stark shift.  The shift on each level, $h\Delta f^{(S)}_{F,i}$, may be written \cite{Berkeland1998}
\begin{equation}
h\Delta f^{(S)}_{F,i}=\tfrac{1}{2}C_{F,m_F}\alpha_{2,J}E_0^2\left(3\cos^2\beta_i-1\right)\left(-\frac{\langle v_i^2\rangle}{2 c^2}\right),
\end{equation}  
where $E_0=m\Omega_\mathrm{rf} c/e$, $h$ is Planck's constant, $c$ is the speed of light, $e$ is the elementary charge, $m$ the mass of the ion, $\Omega_\mathrm{rf}$ is the frequency of the applied trap drive, $\beta_i$ is the angle between the applied magnetic field and the direction of the oscillating field at the ion, and $\alpha_{2,J}$ is the static tensor polarizability for the finestructure level.  The final term is the fractional second order Doppler (SD) shift due to the velocity, $v_i$, of the ion.


\section{Measurements}
The EQ moments are inferred from measurements on the two hyperfine transitions $\ket{^3D_1,7,0}\leftrightarrow\ket{^3D_1,8,0}$ and $\ket{^3D_2,5,0}\leftrightarrow\ket{^3D_2,6,0}$.  In both cases, the differential shift between ions is measured using a servo that treats the inner and outer ions much like a Zeeman pair.  A Ramsey sequence consisting of $1.75\,\mathrm{ms}$ $\pi/2$-pulses separated by 1\,s is used for interrogation.  Measurements alternate either side of the central Ramsey fringe for the outer and inner ions, giving four measurements in total, and  frequencies for the inner and outer pair are updated after $10$ iterations. Since frequency differences between the outer ions are at the mHz level, their signals are combined so that the servo tracks the mean frequency of both.  This cancels small differential shifts between the two outer ions that arise from micromotion and a small magnetic field gradient.  Differences in signal contrast between the two outer ions can degrade this cancellation but, since the differences from EMM and field gradients are small, the resulting error is negligible. 

The magnetic field is aligned to the crystal axis by rotating the magnetic field in the horizontal plane and measuring the differential frequency of the $\ket{^3D_1,7,0}\leftrightarrow\ket{^3D_1,8,0}$ microwave transition between the outer and inner ions.  At each setting, the optical pumping beam is realigned to the field to optimize state preparation.  The experimental sequence begins with $2$\,ms of Doppler cooling followed by optically pumping to $\ket{^3D_1,7,0}$.  After the transition to $\ket{^3D_1,8,0}$ is interrogated, population remaining in $\ket{^3D_1,7,0}$ is double shelved to $\ket{^1S_0,7,\pm1}$ allowing detection of population transferred to $\ket{^3D_1,8,0}$.  Results are plotted in Fig.~\ref{Result}(a), and an Allan deviation for a typical run given in Fig.~\ref{Result}(b).  Each point is determined by several hours of averaging and the error bars are determined by the expected projection noise for the measurement duration. The solid curve is a fit to a quadratic, which has a reduced $\chi^2_\nu$ of $0.9$ and an estimated maximum shift of $1.3750(14)$ Hz.

By virtue of the optical pumping beam alignment, the magnetic field is well aligned to the horizontal plane, but this need not be the same as the crystal axis.  This angle is measured in the same way but at the expense of optical pumping, which limits the range of angles over which measurements can be made.  From such measurements we infer an angular misalignment of $0.3(4)$ degrees relative to the field setting that optimizes optical pumping as used for the measurements in Fig.~\ref{Result}.

For $^3D_2$, only a single measurement of the $\ket{^3D_2,5,0}\leftrightarrow\ket{^3D_2,6,0}$ microwave transition was carried out, with the magnetic field set to the optimum value inferred from the $^3D_1$ measurements (horizontal) and optical pumping (vertical).  The measurement sequence differs only in the state preparation and detection steps.  Specifically, after optical pumping to $\ket{^3D_1,7,0}$, the ions are transferred to $\ket{^3D_2,5,0}$, and the $\ket{^3D_2,5,0}\leftrightarrow\ket{^3D_2,6,0}$ transition interrogated. Remaining population in $\ket{^3D_2,5,0}$ is then transferred back to $\ket{^3D_1,7,0}$ for detection.  A total of 15 hours of data collection gives an estimated differential shift of $0.9021(14)$ Hz and the Allan deviation is given in Fig.~\ref{Result}(c).

\section{Systematics}
\begin{figure*}
  \includegraphics[width=\textwidth]{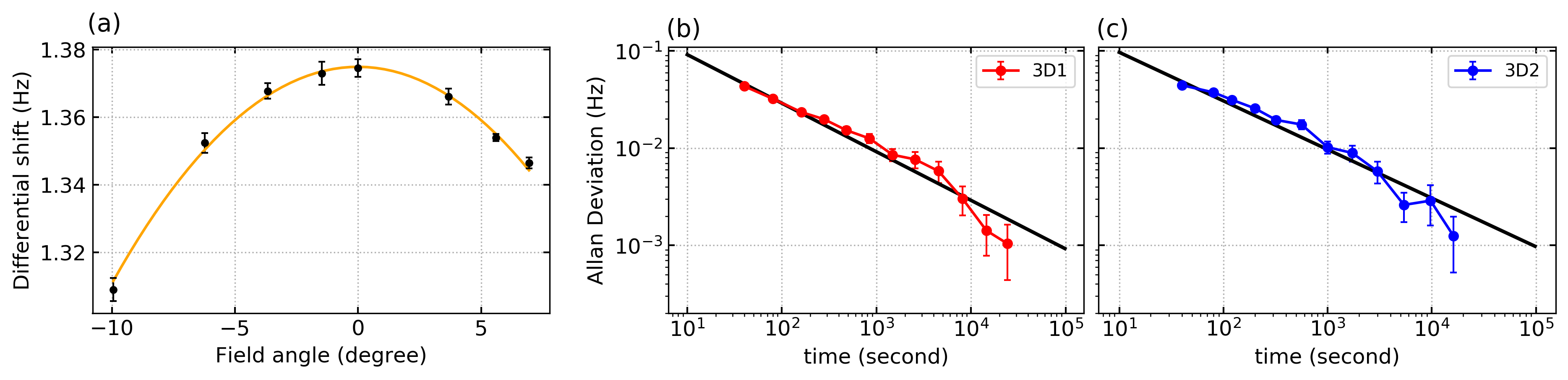}
  \caption{(a) Differential frequency of the $\ket{^3D_1,7,0}\leftrightarrow\ket{^3D_1,8,0}$ microwave transition between the outer and inner ions as a function of magnetic field angle. (b) and (c) are the Allan deviations of the differential microwave frequencies for $^3D_1$ and $^3D_2$, respectively.}
 \label{Result}
\end{figure*}

 \begin{table}
 \caption{Error budget for the quadrupole moment determinations.  Values are given in units of $ea_0^2$ where $a_0$ is the Bohr radius.}
 \label{QuadValues}
\begin{ruledtabular}
\begin{tabular}{l c c c c}
Contribution &  $\Theta(^3D_1)$ & Unc.($10^{-3}$) & $\Theta(^3D_2)$ & Unc.($10^{-3}$) \\
 \hline
 \vspace{-0.4cm}\\
Meas. shift & 0.63791 & 0.65 & 0.85936 & 1.33\\
Trap freq. & - & 0.32 & - & 0.43 \\
Angle ($\theta$) &- & 0.06 & - & 0.09\\
Micromotion & -0.00047 & 0.04 & -0.00050 & 0.05\\
Anharmonicity & -0.00024 &0.12 & -0.00033 & 0.16  \\
\hline
\textbf{Corrected} & 0.63862 & 0.74 & 0.86019 & 1.41 
 \end{tabular}
 \end{ruledtabular}
 \end{table}

 \begin{table}
 \caption{Fractional second-order Doppler shifts for all three ions, along the $x$, $y,$ and $z$ axes as given in Fig.~\ref{ExpSetup}.  Values are given relative to $10^{-18}$.}
 \label{EMM}
\begin{ruledtabular}
\begin{tabular}{c c c c}
 \hspace{0.25cm}  Axis &  \hspace{0.5cm} ion 1 & \hspace{0.25cm} ion 2 & \hspace{0.25cm} ion 3 \hspace{0.5cm} \\
 \hline
 \vspace{-0.4cm}\\
\hspace{0.25cm} $x$ & \hspace{0.25cm} -0.73(2) & \hspace{0.25cm} -0.01(1) & \hspace{0.25cm} -0.56(2) \hspace{0.25cm}\\
\hspace{0.25cm} $y$ & \hspace{0.25cm} -0.27(2) & \hspace{0.25cm} -0.23(2) & \hspace{0.25cm} -0.22(2) \hspace{0.5cm}\\
\hspace{0.25cm} $z$ & \hspace{0.25cm} -1.43(3) & \hspace{0.25cm} -0.04(1)& \hspace{0.25cm} -2.79(6)\hspace{0.1cm}
 \end{tabular}
 \end{ruledtabular}
 \end{table}

The quadrupole shifts inferred from the measurements presented in the previous section are tabulated in Table~\ref{QuadValues} along with the leading systematics and their uncertainties.  As indicated, the uncertainty budget is dominated by the statistical error and otherwise limited by the determination of the trap frequency and possible anharmonicity of the confinement.  The trap frequency was inferred from sideband spectroscopy of the $\ket{^3D_1,7,0}\leftrightarrow\ket{^3D_1,8,0}$ transition and taking the difference between the upper and lower vibrational sidebands to eliminate common mode effects such as ac Stark shifts from the Raman beams.  The uncertainty in the frequency reflects the drift of the trap frequency observed over the duration of the measurements.

Trap anharmonicity results in a varying electric field gradient over the spatial extent of the ion crystal, and hence contributes to the differential frequency measurement.  This was investigated by numerical simulations of the trapping field, which indicated a fractional increase in the electric field gradient of $3.8\times 10^{-4}$ for the outer ions relative to the center.  We have conservatively taken a $50\%$ uncertainty to account for possible imperfections in trap geometry or from stray field contributions, noting that variations in the axial end cap separation by $\pm10\%$ changed the estimate by $\sim 5\%$.
 
The angle $\theta$ accounts for the angular misalignment of the magnetic field with respect to the crystal axis.  For the $^3D_1$ case, this is partially accounted for as the shift is determined from the fitted curve in Fig.~\ref{Result}(a).  Thus the uncertainty in the angle is determined from the vertical direction alone.  As the $^3D_2$ measurement was made by setting the field at the estimated optimum field settings, the uncertainty accounts for both vertical and horizontal directions.

For the measurements described in the previous section, EMM was minimized for a single ion as described in \cite{tan2019suppressing}. Additionally, it was measured for all three ions along the $x$, $y$, and $z$ axes as labeled in Fig.~\ref{ExpSetup}, using the sideband-ratio method \cite{Berkeland1998}.  As in \cite{tan2019suppressing}, EMM in the $xz$ plane is detected using stimulated-Raman transitions between $\ket{^3D_1,7,0}$ and $\ket{^3D_1,8,0}$ with the two sets of 646-nm beams shown in Fig. \ref{ExpSetup}(a).  Along the $y$ direction,  EMM was measured using the 804-nm clock transition, in which the ion is first transferred from $\ket{^3D_1,7,0}$ to $\ket{^1S_0,7,1}$ by the 848-nm laser.  The 804-nm transition to $\ket{^3D_2,5,0}$ is then driven on the rf-sideband, and any remaining population in $\ket{^1S_0,7,1}$ transferred back to $\ket{^3D_1,7,0}$ for detection.  To make full use of the resolution provided by the clock transition, the 804-nm laser power was increased by a factor of $4500$ when driving the sideband.  

The fractional SD shifts inferred from the sideband ratio measurements are tabulated in Table~\ref{EMM}.  In each case, the EMM-induced modulation index is inferred from measured Rabi frequencies when flopping on the sideband and carrier transitions and shifts are then inferred from \cite[Eq.~49]{Berkeland1998}.  Along the axial direction, the difference between the two outer ions is consistent with a small ($\sim 1.4\,\mathrm{\mu m}$) displacement of the center ion and its corresponding non-zero SD shift.  Along the $x$ direction, the larger value for the outer ions is consistent with a $1^\circ$ rotation of the dc confinement field relative to the principle axes determined by the rf field, which might arise from stray fields for example.  The larger value along $y$ for the middle ion arises from the fact that it was minimized using the less sensitive 646\,nm scattering on the rf sideband.  This could be improved using the 804-nm measurements, but it is unnecessary for these experiments.  Indeed the contribution arising from EMM is on the order of the statistical error from the measurement itself.  The uncertainty in the contribution is primarily due to the uncertainty in the tensor polarizability of each transition.

Another possible systematic frequency shift arises from oscillating currents in the electrodes driven by the trap drive at the frequency $\Omega_{\mathrm{rf}}$ \cite{gan2018oscillating}.  The resulting ac Zeeman shift depends on the magnitude and orientation of the ac field relative to the applied dc field and could potentially have a spatial variation along the ion crystal.  This was investigated using the method in \cite{arnold2020precision}.  Measurements were carried out on three $^{138}$Ba$^+$ ions, under three orthogonal orientations of the applied dc field.  These measurements confirmed that the ac magnetic field at each ion was approximately orthogonal to the crystal axis, with very little spatial variation along the axis. Field components at each ion are given in Table~\ref{acMagnetic}.  The most sensitive of the two transitions to the oscillating fields is the $\ket{^3D_2,5,0}\leftrightarrow\ket{^3D_2,6,0}$ transition, for which the inferred differential ac Zeeman shift between outer and inner ions is $-65(47)\,\mathrm{\mu Hz}$. This is well below the statistical precision of the quadrupole determination and hence this systematic has been omitted from table~\ref{QuadValues}.

\begin{table}
 \caption{Amplitudes of the ac magnetic field components for each ion measured with a three-ion crystal of $^{138}$Ba$^+$ ions using the method presented in \cite{arnold2020precision}.  Values are given in $\mu \mathrm{T}.$}
 \label{acMagnetic}
\begin{ruledtabular}
\begin{tabular}{c c c c}
 \hspace{0.25cm} &  \hspace{0.5cm} ion 1 & \hspace{0.25cm} ion 2 & \hspace{0.25cm} ion 3 \hspace{0.5cm} \\
 \hline
 \vspace{-0.4cm}\\
\hspace{0.25cm} $B_\perp$ & \hspace{0.25cm} 0.8153(9) & \hspace{0.25cm} 0.8182(8) & \hspace{0.25cm} 0.8148(10) \hspace{0.25cm}\\
\hspace{0.25cm} $B_\|$ & \hspace{0.25cm} 0.079(11) & \hspace{0.25cm} 0.036(23) & \hspace{0.25cm} 0.083(11) \hspace{0.5cm}\\
 \end{tabular}
 \end{ruledtabular}
 \end{table}

\section{Discussion}
In summary, we have provided measurements of the $^3D_1$ and $^3D_2$ quadrupole moments in $^{176}$Lu$^+$.  Our measurement scheme utilized the differential shifts between ions in a multi-ion crystal for which external electric field gradients largely appear common mode leaving only the well defined Coulomb interaction.  The measurements provided are statistically limited to a precision at the $0.1\%$ level.  In principle this could be improved by increasing the number of ions, which increases the differential shift between ions.  However, this would require better control of micromotion over a larger distance and likely increase the importance of anharmonic terms in the potential.

It should be noted that a quadrupole moment $\Theta(J)$ attributed to a fine-structure level is only an approximation, as the hyperfine interaction leads to corrections to Eq.~\ref{QuadrupoleEq}, which modifies $\Theta(J,F,m_F)$ by $\delta\Theta(J,F,m_F)$.  This has been explored for the special case of a $^3P_0$ level in Al$^+$ and In$^+$ for which the hyperfine interaction leads to a nonzero quadrupole moment \cite{beloy2017hyperfine}.  In lutetium, the leading order correction is given by

\begin{multline}
\delta\Theta(J,F,m_F)=2(-1)^{F-m_F}(2F+1)\\
\begin{pmatrix} F & 2 & F\\ -m_F & 0 & m_F\end{pmatrix}\begin{pmatrix} I & 1 & I\\ -I & 0 & I\end{pmatrix}^{-1}\\
\times \sum_{J'} \begin{Bmatrix} F & F & 2\\ J & J' & I\end{Bmatrix} \begin{Bmatrix}F & J' & I \\1 & I & J\end{Bmatrix}\beta^Q_{J,J'},
\end{multline}
where 
\begin{equation}
\beta^Q_{J,J'}=\frac{\mu_I}{\mu_N} \frac{\matrixelred{\gamma J}{\Theta^{(2)}}{\gamma' J'}\matrixelred{\gamma' J'}{\vb*{T}^e_{1}}{\gamma J}}{E_{\gamma,J}-E_{\gamma',J'}},
\label{Eq:betaQ}
\end{equation}
with $\mu_I$ the magnetic dipole moment of the nucleus, $\mu_N$ the nuclear magneton, and $\vb*{T}^e_{1}$ the rank 1 electronic tensor from the hyperfine interaction \cite{beloy2008hyperfine}.  

From matrix elements given in \cite{paez2016atomic,kaewuam2019spectroscopy}, hyperfine-mediated corrections to the EQ moment of $^3D_1$ states are dominated by contributions from $^3D_2$, for which $\beta^Q_{1,2}=-0.014 e a_0^2$.  The corresponding fractional contribution to the estimate of $\Theta({}^3D_1)$ is approximately $1.3\times10^{-3}$, which is comparable to the uncertainty given in table~\ref{QuadValues}.  Hence improved accuracy in determining $\Theta({}^3D_1)$ would have to account for these terms.  Moreover, the corrections to the EQ moment of a given state can be as large as $0.5\%$ of $C_{F,m_F} \Theta({}^3D_1)$ .  Similar considerations also apply to $^3D_2$, but currently available matrix elements are insufficient to make more quantitative statements.  Consequently we tentatively give 
\begin{align}
\Theta({}^3D_1)&=0.63862(74)\:ea_0^2,\\
\Theta({}^3D_2)&=0.8602(14)\:ea_0^2,
\end{align}
as the quadrupole moments of $^3D_1$ and $^3D_2$, respectively, with the reservation that hyperfine-mediated terms have not been accounted for.

CI+all-order calculations \cite{porsev2018clock} yielded 0.653\,$ea_0^2$ and 0.885\,$ea_0^2$ for the quadrupole moments of the $^3D_1$  and $^3D_2$ states, respectively, taking into account the factor of 2 difference in the definition of the quadrupole moment in \cite{porsev2018clock} and \cite{Itano2000}, which is used here. The estimated accuracy of these values is 1.5\% and 3.1\%, respectively. These uncertainties are obtained as the differences of the CI+all-order and another calculation, where CI is combined with the many-body perturbation theory. Such a method is commonly used in the evaluation of the uncertainties of the CI+all-order values for various properties \cite{MasHoeWan19}. The theory values agree with the experimental results to within 2.2\% and 2.9\%, for the $^3D_1$  and $^3D_2$ states, respectively, which is supportive of the method used to estimate the theoretical uncertainties.  Consideration of the much smaller hyperfine-mediated terms does not affect this conclusion.

In principle, $\beta^Q_{J,J'}$ could be determined by comparing measurements over multiple microwave transitions.  However, there is currently insufficient accuracy for this to provide a meaningful estimate of $\beta^Q_{J,J'}$.  An alternative possibility, is to note that uncertainty in the theoretical value is primarily due to the matrix element $\matrixelred{\gamma' J'}{\vb*{T}^e_{1}}{\gamma J}$.  This same matrix element appears in the hyperfine-mediated corrections to the electronic $g_F$-factors.  Thus, accurate $g_F$-factor measurements may provide an experimental determination of $\matrixelred{\gamma' J'}{\vb*{T}^e_{1}}{\gamma J}$.  A more detailed analysis of hyperfine-mediated effects and $g$-factor measurements is currently in progress.
\begin{acknowledgements}
This work is supported by the National Research Foundation, Prime Ministers Office, Singapore and the Ministry of Education, Singapore under the Research Centres of Excellence programme.  It is also supported by the National Research Foundation, Singapore, hosted by National University of Singapore under the Quantum Engineering Programme (Grant Award QEP-P5).  M.S.S. acknowledges the sponsorship of ONR Grants No. N00014-17-1-2252 and N00014-20-1-2513.
\end{acknowledgements}
\bibliography{LuQuadrupole}
\bibliographystyle{unsrt}
\end{document}